\documentstyle[11pt]{article}
\begin{document}
\title{\Large \bf Solitons on Noncommutative Orbifold $ T^2/Z_N $}
\author{ Bo-yu Hou$$ \thanks{Email:byhou@phy.nwu.edu.cn}, 
\hspace{5mm}
        Kang-jie Shi$$ \thanks{Email:kjshi@phy.nwu.edu.cn},
                         \hspace{5mm}
        Zhan-ying Yang$$ \thanks{Email:yzy@phy.nwu.edu.cn}\\[3mm]
        Institute of Modern Physics, Northwest University,\\
        Xi'an, 710069, P. R. China}                          
\maketitle
\vspace{1cm}
\begin{abstract}
Following the construction of the projection operators on $ T^2 $ 
presented by Gopakumar, Headrick and Spradlin, we construct a set of 
projection operators on the integral noncommutative orbifold $ 
T^2/G (G=Z_N, N=2, 3, 4, 6)$ which correspond to a set of solitons on 
$T^2/Z_N$ in noncommutative field theory. In this way, we derive an 
explicit form of projector on $T^2/Z_6$ as an example. We also construct 
a complete set of projectors on $T^2/Z_N$ by series expansions for integral 
case. 
\end{abstract}

\medskip
\medskip

\medskip
\section{Introduction}
\par Noncommutative geometry is originally an interesting topic in 
mathematics[1][2][3].
 In the last few years, noncommutative field theories have renewed the 
physist's interest primarily due to 
the discovery that non-commutative gauge theories naturally arise from the 
low energy dynamics of D-branes in the presence of a background $ B$ field 
and as various limits of M-theory compactification[4, 5, 6].
Quantum field theory on a noncommutative space is useful to understand 
various physical phenomena, such as string behaviors and D-brane dynamics.
They also appear as theories describing the behavior of the  
electron gas in the presence of a strong, external magnetic field, the 
quantum Hall effect[7]. Recently Susskind and Hu, Zhang[8] proposed that 
noncommutative Chern-Simons theory on the plane may provide a description 
of (fractionally filled) quantum Hall fluid. Being nonlocal,
noncommutative field theory may help to understand nonlocality at short 
distant in quantum gravity. 

After the connection between string theory and noncommutative 
field theories was unraveled, the study of solitons in 
noncommutative space have attracted much attention[9, 10, 11, 12, 13]. 
Soliton 
solutions in field theory and string theory often shed 
light on the nonperturbative and strong coupling behavior of the theory,
thus it is interesting to investigate these solutions in noncommutative 
fields theories. 
Gopakumar, Minwalla and Strominger found that 
soliton solution 
of noncommutative flat space can be exactly given in terms of 
projection operators[12]. Harvey et al set up a new method to investigate 
the soliton solution and M. Hamanaka and S. Terashima generalised this 
"Solution Generating Technique" to BPS monopole 
in 3+1 dimension[13]. Martinec and Moore
discussed how D-branes on orbifolds fit naturally into the 
algebraic framework as described by projection operators[10]. 
Thus the study of projection operators in various noncommutative spaces 
are important in string field theory.
Rieffel has presented a general 
formula for the projection operators on noncommutative torus[14]. 
Boca further described 
the projection operators on orbifold $T^2/G$ and discussed 
the relation between their trace and the commutator $q$ of the 
operators $U$ and $V$ on noncommutative torus$(UV=VUe^{2i\pi 
q})$. He proved the existence of nontrivial projection 
operators and explicitly presented an 
example with trace $1/q$ for $q$ an integer in $Z_4$ case [15]. Boca 
expressed the 
projection operators in terms of the $\theta$ 
function depend respectively on $U$ and $V$. 
Konechny, Schwartz and Walters [16, 17] have also given some $Z_2, Z_4$ 
invariantprojection operators. Martinec and Moore  
pointed out that no explicit expressions for the projection
operators in $Z_3, Z_6$ case have been found.  Gopakumar et al 
[9] succeeded in constructing the projection operator on 
noncommutative integral torus with generic $\tau$. Here the integer $A$ 
measures the quantuns of magnetic flux passing through the torus, its 
ratio with the area servers as the noncommutative parameter. We find that 
if the vacuum state $|0>$ in their paper is replaced by any
state vector $|\phi>$, their construction still works. We notice that 
if the state vector $|\phi>$ has some symmetries, the 
operators are just the projection operators on orbifold $T^2/G$ ($G=Z_N$ 
is a symmetry group). We then give a set of projection operators in 
$T^2/Z_N$. As an example, an explicit projection operator in $T^2/Z_6$ is 
obtained by this approach.
   
This paper is organized as following. We introduce the noncommutative 
orbifold $T^2/Z_N$ in section 2 and in next section we review the 
construction 
proposed by Gopakumar et al on the integral torus $T^2$. We show how this 
approach can be used
to construct the projection operators which is 
invariant under the transformation group $Z_N$
in section 4. In section 
5, we presented an explicit form of projector on $T^2/Z_6$ as an example, 
using theta functions with $\hat{y_1}$ and $\hat{y_2}$ as variables. 
In section 6, we provided a complete set of projectors on 
$T^2/Z_N$ by series expansions for integral case.
\section {Noncommutative Orbifold $T^2/Z_N$}
 In this section, we introduce operators on the noncommutative 
orbifold $T^2/Z_N$. First we introduce two 
operators $\hat{y_1}$ and $\hat{y_2}$ on noncommutative $R^2$ which 
satisfy the following commutation relation:
\begin{equation}
[\hat{y_1},\hat{y_2}]=i.
\end{equation}
 Define operators
\begin{equation}
U_1=e^{-il\hat{y_2}},~~~~~~~~~~
U_2=e^{il(\tau_2\hat{y_1}-\tau_1\hat{y_2})},
\end{equation}
where $l, \tau_1, \tau_2$ are all real numbers and $l, \tau_2>0$. All 
operators 
on $R^2$ which commute with $U_1$ and $U_2$ constitute the operators 
defined on 
noncommutative torus $T^2$.  This torus is formed as manifold which 
identify
two points $(\hat{y_1}, \hat{y_2}) \sim (\hat{y_1}, \hat{y_2}) +{\bf 
r}$ with ${\bf r}=m{\bf l_1} + n{\bf l_2}$ on noncommutative 
plane $R^2$, where ${\bf l_1}=(l, 0), {\bf l_2}=(l\tau_1, l\tau_2)$. 
Thus we have
\begin{eqnarray}
U_1^{-1}\hat{y_1}U_1&=& \hat{y_1} +l,~~~~~~~U_2^{-1}\hat{y_1}U_2= 
\hat{y_1} +l\tau_1,\\
U_1^{-1}\hat{y_2}U_1&=& \hat{y_2},~~~~~~~~~~~U_2^{-1}\hat{y_2}U_2= 
\hat{y_2}
 +l\tau_2.\nonumber
\end{eqnarray}
The operators $U_1$ and $U_2$ are two different wrapping operators 
around the noncommutative torus and their commutation relation is 
$U_1U_2=U_2U_1e^{-2\pi 
i\frac{l^2\tau_2}{2\pi}}$. When $A=\frac{l^2\tau_2}{2\pi}$ is an integer 
, we call the torus integral. Next we introduce a 
linear transformation $R$, which gives
\begin{equation}
R^{-1}\hat{y_1}R=a\hat{y_1}+ 
b\hat{y_2},~~~~~~~R^{-1}\hat{y_2}R=c\hat{y_1}+ d\hat{y_2}.
\end{equation}
 Setting $U_1=U,U_2=V$, Under $R$, if $U$ and $V$ change as  
[10]\footnote[1]{These relations may include some phase factors. It will 
not 
change the derivation in the text, see equation (10).} 
\begin{eqnarray}
&& Z_2: ~~~~ U\rightarrow U^{-1},~~~~~ V\rightarrow V^{-1},\nonumber\\
&& Z_3: ~~~~ U\rightarrow V,     ~~~~~ V\rightarrow 
U^{-1}V^{-1},\nonumber\\
&& Z_4: ~~~~ U\rightarrow V,     ~~~~~ V\rightarrow U^{-1},\nonumber\\
&& Z_6: ~~~~ U\rightarrow V,     ~~~~~ V\rightarrow U^{-1}V,
\end{eqnarray}
then we refer $R$ as a $Z_N$ symmetry rotation of the torus $T^2$. The 
operators on noncommutative orbifold $T^2/Z_N$ are the operators of $T^2$
which are invariant under transformation $R$. 

If we define operators $\hat{y'_1}$ and $\hat{y'_2}$ as 
\begin{eqnarray}
\hat{y_1}&=&a\hat{y'_1}+b\hat{y'_2},~~~~~~~\hat{y_2}=\frac{1}{a}\hat{y'_2},\\
a&=&\sqrt{\frac{\tau'_2}{\tau_2}}, 
~~~b=\frac{-\tau'_1+\tau_1}{\sqrt{\tau_2\tau'_2}},~~~l'=\frac{l}{a},\nonumber
\end{eqnarray}
then we get
\begin{eqnarray}
[\hat{y'_1},\hat{y'_2}]&=&i, ~~~~~~~U_1=e^{-il'\hat{y'_2}},\\
U_2&=&e^{il'(\tau'_2\hat{y'_1}-\tau'_1\hat{y'_2})}.\nonumber
\end{eqnarray}
From the above result, we notice that the noncommutative 
torus is invariant after taking a suitable module parameter 
$\tau=\tau_1+i\tau_2$. 
 Now we consider the rotations in symmetric 
orbifolds $T^2/Z_N (Z_N=Z_2, Z_3, Z_4, Z_6)$. Let $\tau=e^{\frac{2\pi 
i}{N}}=e^{i\theta}$,
then the transformation 
\begin{equation}
R^{-1}\hat{y_1}R=\cos \theta \hat{y_1}+\sin \theta \hat{y_2},
~~~~~~~R^{-1}\hat{y_2}R=\cos \theta \hat{y_2}- \sin \theta \hat{y_1}.
\end{equation} 
will give corresponding transformation of the operators $U_1$ and $U_2$ 
as in equation (5). Such $R$ can be realized by
\begin{equation}
 R=e^{-i\theta
\frac{\hat{y_1}^2+\hat{y_2}^2}{2}+i\frac{\theta}{2}}.
\end{equation} 
Next we take $Z_6$ and $Z_4$ as examples.\\
(1) $\theta=\frac{\pi}{3}$ 
\begin{eqnarray}
\tau_1&=&\frac{1}{2}, ~~~~~\tau_2=\frac{\sqrt{3}}{2},~~N=6,\nonumber\\
U_1&=&e^{-il\hat{y_2}},~~~~~~
U_2=e^{il(\frac{\sqrt{3}}{2}\hat{y_1}-\frac{1}{2}\hat{y_2})},\nonumber\\
R^{-1}U_1 R&=&U_2,~~~~~~~~~R^{-1}U_2 R=e^{-\pi iA}U_1^{-1}U_2.
\end{eqnarray}
From the above result, we find that the lattice remain invariant under 
rotation.\\
(2)$\theta=\frac{\pi}{2}$ 
\begin{eqnarray}
\tau_1&=&0, ~~~~~\tau_2=1,~~~~~N=4,\nonumber\\
U_1&=&e^{-il\hat{y_2}},~~~~~~U_2=e^{il\hat{y_1}},\nonumber\\
R^{-1}U_1 R&=&U_2,~~~~~~~~~R^{-1}U_2 R=U_1^{-1}.
\end{eqnarray}      
 This shows that the whole lattice remain invariant. We can realize the 
operators $\hat{y_1}$ and $\hat{y_2}$ as the operators
in Fock space. Introducing 
\begin{equation}
a=\frac{\hat{y_2}-i\hat{y_1}}{\sqrt{2}},~~~~~~
a^{+}=\frac{\hat{y_2}+i\hat{y_1}}{\sqrt{2}}
\end{equation}
and we have $[a,a^{+}]=1$. The rotation $R$ can be expressed by $a, a^{+}$ 
via
\begin{equation} 
R=e^{-i\theta a^{+}a}.
\end{equation} 
\section{ Review GHS Construction for Soliton}
In this section, we review the results in paper [9].
A noncommutative space $R^2$ has been orbifolded to a torus $T^2$ with
double periods $l$ and $\tau l$. The generators are
\begin{equation}
U_1=e^{-il\hat{y_2}},~~~~~~~~
U_2=e^{il(\tau_2\hat{y_1}-\tau_1\hat{y_2})}
\end{equation}
where $[\hat{y_1},\hat{y_2}]=i$, here we just consider the case when
$A=\frac{\tau_2 l^2}{2\pi}$ is an integer. Introduce a state
vector
\begin{equation}
|\psi>=\sum_{j_1,j_2}C_{j_1,j_2} U_1^{j_1}U_2^{j_2}|\Omega> (j_1,j_2 \in 
Z)
\end{equation}
that satisfies
\begin{equation}
<\psi|U_1^{j_1}U_2^{j_2}|\psi>=\delta_{j_1, 0}\delta_{j_2, 0}.
\end{equation}
The state $|\Omega>$ will be specified later. Then a projection operator 
on $T^2$ can be constructed as
\begin{equation}
P=\sum_{j_1,j_2} U_1^{j_1}U_2^{j_2}|\psi><\psi|U_2^{-j_2}U_1^{-j_1}.
\end{equation}
The power series of $\hat{y_1}$ and $\hat{y_2}$ can be made up 
of the power series of $a$ and $a^{+}$. Moreover the formula 
$|0><0|=:e^{-a^{+}a}:$ indicates that any $|\psi><\psi|$ can be 
constituted
by the power series of $a$ and $a^{+}$. The projection operator is 
therefore 
spanned by the operators $\hat{y_1}$ and $\hat{y_2}$.
It is easy to check $P^2=P$ and $U_{i}^{-1}P U_i=P$. So $P$ is 
an projection operator on noncommutative $T^2$. The $kq$ 
representation[18][19] provides a basis of 
common eigenstate of $U_1$ and $U_2$:
\begin{equation}
|k,q>=\sqrt{\frac{l}{2\pi}}e^{-i\tau_1\hat{y_2}^2/2\tau_2}\sum_{j}
e^{ijkl}|q+jl>
\end{equation}
where the ket on the right is a $\hat{y_1}$ eigenstate. We have
\begin{eqnarray}
U_1|k,q>&=&e^{-ilk}|k,q>,~~~~~~~U_2|k,q>=e^{il\tau_2 q}|k,q>,\nonumber\\
id&=&\int_{a}^{\frac{2\pi}{l}+a}dk\int_{b}^{l+b}dq |k,q><k,q|
\end{eqnarray}
where $a$ and $b$ are real numbers and 
$|k,q>=|k+\frac{2\pi}{l},q>=e^{ilk}|k,q+l>$.
 In terms of wave functions in the $kq$ representation, $|\psi>$ becomes
\begin{equation}
C_{\psi}(k,q)\equiv 
<k,q|\psi>=\sum_{j_1,j_2}C_{j_1,j_2}e^{-ij_1 lk +ij_2 l\tau_2 
q}<k,q|\Omega>=\tilde{c}(k,q)C_0(k,q)
\end{equation}
where $\tilde{c}(k,q)=\sum_{j_1,j_2}C_{j_1,j_2}e^{-ij_1 lk+ij_2 l\tau_2 
q},
C_0(k,q)=<k,q|\Omega>$. Note that  $\tilde{c}(k,q)$ is doubly periodic: 
$$\tilde{c}(k+\frac{2\pi}{l},q)= \tilde{c}(k,q+\frac{l}{A})= 
\tilde{c}(k,q).$$ The 
orthonormality condition (16) becomes
\begin{equation}
\delta_{j_1,0}\delta_{j_2,0}=\int_{0}^{\frac{2\pi}{l}}dk\int_{0}^{l}dq
e^{-ij_1 lk+ij_2 l\tau_2 q}|\tilde{c}(k,q)|^2|C_0(k,q)|^2.
\end{equation}
The coefficient $C_{j_1 j_2}$ can be obtained if and only if 
$\tilde{c}(k,q)$
is double periodic function with periods  $2\pi/l$ and $l/A$ for $k$ and 
$q$ respectively. So we rewrite the above equation as
\begin{equation}
\delta_{j_1 0}\delta_{j_2 0}=\int_{0}^{\frac{2\pi}{l}}dk\int_{0}^{l/A}dq
e^{-ij_1 lk+ij_2 l\tau_2 
q}|\tilde{c}(k,q)|^2\sum_{n=0}^{A-1}|C_0(k,q+\frac{ln}{A})|^2.
\end{equation} 
This hold for any $j_1$ and $j_2$ if and only if $
|\tilde{c}(k,q)|^2\sum_{n=0}^{A-1}|C_0(k,q+\frac{ln}{A})|^2=\frac{A}{2\pi}$.
Then we have
\begin{equation}
|\tilde{c}(k,q)|=\frac{1}{\sqrt{\frac{2\pi}{A}\sum_{n=0}^{A-1}
|C_0(k,q+\frac{ln}{A})|^2}}.
\end{equation}
Setting $e^{i\beta}$ as phase factor of $\tilde{c}(k,q)$, we have
\begin{equation}
C_{\psi}(k,q)=\frac{C_0(k,q)e^{i\beta}}{\sqrt{\frac{2\pi}{A}\sum_{n=0}^{A-1}
|C_0(k,q+\frac{ln}{A})|^2}}.
\end{equation} 
\section{The Projection Operator on $T^2/Z_N$}
 In the last section, we reviewed how to construct the projection 
operators on noncommutative torus. In this section, we will discuss how to 
construct the projection operator on the noncommutative orbifold $T^2/Z_N$ 
following the result of the last section. Recall the 
projection operator 
\begin{equation}
P=\sum_{j_1,j_2} U_1^{j_1}U_2^{j_2}|\psi><\psi|U_2^{-j_2}U_1^{-j_1}
\end{equation}    
and transform it by rotation $R$
\begin{equation}
R^{-1}PR=\sum_{j_1,j_2} 
(U'_1)^{j_1}(U'_2)^{j_2}R^{-1}|\psi><\psi|R 
(U'_2)^{-j_2}(U'_1)^{-j_1}
\end{equation} 
where $U'_i=R^{-1}U_i R$. Considering the transformation group $G=Z_N$, we 
get
\begin{equation}
R^{-1}PR=\sum_{j'_1,j'_2}
U_1^{j'_1}U_2^{j'_2}R^{-1}|\psi><\psi|R U_2^{-j'_2}U_1^{-j'_1}
\end{equation}
where $j'_1=-j_1, j'_2=-j_2$ for $Z_2$ case, $j'_1=-j_2, j'_2=j_1-j_2$ for 
$Z_3$ case, $j'_1=-j_2, j'_2=j_1$ for $Z_4$ case and $j'_1=-j_2, 
 j'_2=j_1+j_2$ 
for $Z_6$ case. Then we can obtain $R^{-1}PR=P$ as long as 
\begin{equation}
R|\psi>=e^{i\alpha}|\psi>.
\end{equation}
We can show that this can be satisfied if $R|\Omega>=e^{i\alpha}|\Omega>$.
In the next step, we take 
$G=Z_6$ as an example to prove this (it is easy to 
generalize this to other cases).
Assume
\begin{equation}
|\psi>=\sum_{j_1,j_2}C_{j_1,j_2} U_1^{j_1}U_2^{j_2}|\Omega> (j_1,j_2 \in
Z)
\end{equation}
satisfies
\begin{equation}
<\psi|U_1^{j_1}U_2^{j_2}|\psi>=\delta_{j_1, 0}\delta_{j_2, 0}.
\end{equation}
Setting $ R|\Omega>=e^{i\alpha}|\Omega>$, we have
\begin{eqnarray}
R|\psi>&=&\sum_{j_1,j_2}C_{j_1,j_2} 
U_1^{-j_2}U_2^{j_1+j_2}R|\Omega>\nonumber\\
&=&\sum_{j_1,j_2}C'_{j_1,j_2} U_1^{j_1}U_2^{j_2}|\Omega> 
\end{eqnarray}    
where
\begin{equation}
C'_{j_1, j_2}=C_{j_1+j_2, -j_1}e^{i\alpha}
\end{equation}
 and 
\begin{eqnarray}
<\psi|R^{-1}U_1^{j_1}U_2^{j_2}R|\psi>&=&<\psi|U_2^{j_1}(U_1^{-j_2}U_2^{j_2})
|\psi>\nonumber\\
&=&\delta_{j_1+j_2, 0}\delta_{-j_2, 0}=
\delta_{j_1, 0}\delta_{j_2, 0}.
\end{eqnarray}  
Adding a condition $C^{*}_{j_1, j_2}=C_{-j_1,-j_2} 
e^{-2i\beta}$, we have 
\begin{equation}
\tilde{c}^*(k,q)=\tilde{c}(k,q)e^{-2i\beta}. 
\end{equation}
The 
unique solution satisfying the above equations is 
\begin{equation}
\tilde{c}(k,q)=\frac{e^{i\beta}}{\sqrt{\frac{2\pi}{A}\sum_{n=0}^{A-1}
|C_0(k,q+\frac{ln}{A})|^2}}.
\end{equation}
In brief, The $C_{j_1, j_2}$ which satisfies the condition(29, 30, 34 ) is 
uniquely determined as demonstrated in section 3. 
On the other hand, from equation (32) we find  
\begin{equation}
(C'_{j_1, j_2})^{*}=C'_{-j_1,-j_2}e^{-2i(\alpha+\beta)},
\end{equation}
 giving
\begin{equation}
\tilde{c'}^{*}(k,q)=\tilde{c'}(k,q)e^{-2i(\alpha+\beta)}.
\end{equation}
Equations (31)(33) and (37)also uniquely determine $\tilde{c'}(k,q)$ as
\begin{equation}
\tilde{c'}(k,q)=\frac{e^{i(\alpha+\beta)}}{\sqrt{\frac{2\pi}{A}
\sum_{n=0}^{A-1}
|C_0(k,q+\frac{ln}{A})|^2}}=\tilde{c}(k,q)e^{i\alpha}
\end{equation}
for $R|\psi>$. We then have 
$|\psi'>=R|\psi>=e^{i\alpha}|\psi>$.
In conclusion, if the vector $|\Omega>$ satisfies $ 
R|\Omega>=e^{i\alpha}|\Omega>$, the state $|\psi>$ will satisfy 
equation (28) and the projection operator by GHS construction will be a 
projection operator on noncommutative orbifold $T^2/G$. 

In the above discussion, we know that the crucial point is that
the state vector $|\Omega>$ must be invariant under the rotation
$R$, namely $R|\Omega>=e^{i\alpha}|\Omega>$. Now we show how
to construct such state vector. The operator for
rotation is
\begin{equation}
R=e^{-i\theta a^{+}a}.
\end{equation}
We can set
\begin{equation}
|\Omega>=\sum_{j=0}^{N-1} R^{j}e^{\frac{2\pi i}{N}js}|\phi>
\end{equation}
where $|\phi>$ is an arbitrary state vector, $R=e^{-\frac{2\pi i}{N}a^+
a}, s=0, \cdots, N-1$. Since
\begin{eqnarray}
R^N|\phi>&=&e^{-2\pi i a^+a} (\sum_{n}c_{n}(a^{+})^n)|0>\nonumber\\
&=&\sum_{n}e^{-2\pi in}c_{n}(a^{+})^n|0>\nonumber\\
&=&|\phi>,              
\end{eqnarray}
thus
\begin{eqnarray}
R|\Omega>&=&\sum_{j=0}^{N-1} R^{j+1}e^{i\frac{2\pi}{N}js}|\phi>\nonumber\\
&=&e^{-is\frac{2\pi}{N}}|\Omega>\nonumber\\
&=&e^{i\alpha}|\Omega>.
\end{eqnarray}
If we obtain the expression
for $C_0(k,q)$, it is easy to write the expression for the field
configuration for the
projection operator by equation (24) [9] or get the Fourier expansion by
equations (55)(56).
In next step, we take an example to show how to construct $C_0(k,q)=
<k,q|\Omega>$.
Introduce coherent states
\begin{eqnarray}
|z>&=&e^{-\frac{1}{2}z\bar{z}}e^{a^+z}|0>
\end{eqnarray}
where $z=x+iy, \bar{z}=x-iy$, which satisfies  
\begin{equation}
\frac{1}{\pi}\int_{-\infty}^{\infty}dxdy |z><z|=identity.
\end{equation}
Thus from equation(39), we get
\begin{eqnarray}
R|z>&=&e^{-\frac{1}{2}z\bar{z}}e^{a^+\omega z}|0>\nonumber\\
&=&|\omega z>
\end{eqnarray}
where $\omega=e^{-i\frac{2\pi}{N}}$,
we can employ $<y_2|0>=\frac{1}{\pi^{1/4}}e^{-y_2^2/2}$, $|y_2>$ is  
the eigenstate of the operator $\hat{y_2}$, to obtain
\begin{equation}
<y_2|z>=\frac{1}{\pi^{1/4}}e^{-z^2/2-z\bar{z}/2}e^{-
y_2^2/2+\sqrt{2}zy_2}.
\end{equation}
We have
\begin{eqnarray}
<k,q|z>&=&\int<k,q|y_2><y_2|z>dy_2\nonumber\\
&=&\frac{1}{\sqrt{l}\pi^{1/4}}\theta
\left[   
\begin{array}{l}
0\\
0
\end{array}
\right]
(\frac{q+\frac{\tau}{\tau_2}k-i\sqrt{2}z}{l},\frac{\tau}{A})
e^{-\frac{\tau}{2i\tau_2}k^2+ikq+\sqrt{2}kz-(z^2+z\bar{z})/2}.
\end{eqnarray}
Letting
\begin{equation} 
|\phi>=\frac{1}{\pi}\int_{-\infty}^{\infty}dxdy|z><z|\phi>
=\int_{-\infty}^{\infty}dxdy F(z)|z>,
\end{equation}
we can get
\begin{eqnarray}
C_0(k,q)&=& <k,q|\Omega>=\frac{1}{\pi} \int <k,q|\sum_{j=0}^{N-1}
R^{j}e^{i\frac{2\pi}{N}js}|z><z|\phi>dxdy\nonumber\\
&=&\int [\sum_{j=0}^{N-1}e^{i\frac{2\pi}{N}js}<k,q|\omega^jz>]F(z)dxdy
\end{eqnarray}
where $F(z)$ is an arbitrary function.
After obtaining $C_{0}(k,q)$, we can then compute $<k,q|\psi>$ by 
equation (24) and the Fourier coefficient for the
projection operator $P$ on noncommutative
orbifold $T^2/Z_N$ by equation (56).

\section{An Example for $T^2/Z_6$ case}

In this section, we give an example for $T^2/Z_6$ case. We first review 
the Weyl-Moyal transformation on torus and then
present the explicit expression for the 
projection operator by the Fourier series of the operator $\hat{y_1}$ and 
$\hat{y_2}$. 
Define
\begin{equation}
A(\hat{p})=\sum_{j_1,j_2}U_1^{j_1}U_2^{j_2}b(\hat{p})U_2^{-j_2}U_1^{-j_1} 
(j_1,j_2 \in Z)
\end{equation}   
where $U_1=e^{is_2 \hat{p_1}}, U_2=e^{is_1 \hat{p_2}}$ and $\hat{p_1}, 
\hat{p_2}$ are linear combinations of $\hat{y_1}$ and $\hat{y_2}$, $ 
[\hat{p_1},\hat{p_2}]=i$. It is easy to see that 
$U_i^{-1}A(\hat{p})U_i=A(\hat{p})$, namely $A(\hat{p})$ is an 
operator on noncommutative torus $T^2$.
The field configuration for $A(\hat{p})$ is
\begin{eqnarray}
\Phi_{A}(p)&=&\frac{(2\pi)^2}{s_1s_2}
\sum_{mn} tr\{ e^{2\pi i\lbrack(\hat{p_1}-p_1)\frac{m}{s_1}+
(\hat{p_2}-p_2)\frac{n}{s_2}\rbrack}b(\hat{p})\}. 
\end{eqnarray}
We can also reobtain $A(\hat{p})$ by the Weyl-Moyal transformation from 
$\Phi_{A}(p)$,
\begin{equation}
A(\hat{p})=\sum_{mn}\frac{1}{2\pi s_1 
s_2}\int_{0}^{s_1}dp_1\int_{0}^{s_2}dp_2 \Phi_{A}(p)
e^{2\pi i[(\hat{p_1}-p_1)\frac{m}{s_1}+(\hat{p_2}-p_2)\frac{n}{s_2}]}.
\end{equation}
We now set $\hat{p_1}=-\hat{y_2}, 
\hat{p_2}=\hat{y_1}-\frac{\tau_1}{\tau_2}\hat{y_2}, s_1=l\tau_2, s_2=l,
b(\hat{p})=|\psi><\psi|$, and  have  
\begin{equation}
U_1=e^{-il\hat{y_2}},~~~~~~~~
U_2=e^{il(\tau_2\hat{y_1}-\tau_1\hat{y_2})}.
\end{equation}  
Then the operator $A(\hat{p})$ becomes the projection operator $P$.
The field configuration for the projection operator $P$ is
\begin{equation}
\Phi_{p}(y)=\frac{(2\pi)^2}{l^2\tau_2}
\sum_{j_1j_2}<\psi|e^{\frac{2\pi i}{l}(j_1(\hat{y_1}-y_1)+
\frac{j_2-\tau_1j_1}{\tau_2}(\hat{y_2}-y_2)}|\psi>.
\end{equation}
The Fourier expansion for the projection operator is obtained by 
Weyl-Moyal transformation (52),
\begin{eqnarray}
P&=& 
\sum_{j_1j_2}D_{j_1j_2}e^{-\frac{2\pi 
i}{l}(j_1\hat{y_1}+\frac{j_2-\tau_1j_1}
{\tau_2}\hat{y_2})},\\
D_{j_1j_2}&=&\frac{1}{A}<\psi|e^{\frac{2\pi i}{l}(j_1\hat{y_1}+
\frac{j_2-\tau_1j_1}{\tau_2}\hat{y_2})}|\psi>\nonumber\\
&=&\frac{1}{A}\int_{0}^{\frac{2\pi}{l}}dk\int_{0}^{l}dq 
<k,q|\psi><\psi|k,q-\frac{ls}{A}>
e^{2\pi ij_1(q/l-s/A)}e^{imlk}e^{\pi ij_1j_2/A}
\end{eqnarray}
where $j_2=mA+s, s=0, \cdots, A-1$. 

Rewrite the above equations as
\begin{eqnarray}
P&=&\sum_{j_1j_2}D_{j_1j_2}e^{-2\pi ij_2 \frac{\hat{y_2}}{l\tau_2}}
e^{-2\pi ij_1 \frac{\hat{y_1}-\frac{\tau_1}{\tau_2}\hat{y_2}}
{l}}e^{-\pi i j_1j_2/A}\nonumber\\
&=&\sum_s(e^{-2\pi i \frac{\hat{y_2}}{l\tau_2}})^s
\sum_{j_1 m}(e^{-2\pi iA \frac{\hat{y_2}}{l\tau_2}})^m (e^{-2\pi 
i \frac{\hat{y_1}-\frac{\tau_1}{\tau_2}\hat{y_2}}{l}})^{j_1} 
D_{j_1 ms}\nonumber\\
&=&\sum_s(u_1)^s\sum_{j_1 m}D_{j_1 m s}(u_1^A)^m(u_2)^{j_1}
\end{eqnarray}
where $u_1=e^{-2\pi i \frac{\hat{y_2}}{l\tau_2}}, u_2=e^{-2\pi i 
\frac{\hat{y_1}-\frac{\tau_1}{\tau_2}\hat{y_2}}{l}}$ and
\begin{eqnarray}
D_{j_1 m s}&=&D_{j_1 j_2}e^{-\pi ij_1j_2/A}|_{j_2=mA+s}\nonumber\\
&=&\frac{1}{A}\int_{0}^{\frac{2\pi}{l}}dk\int_{0}^{l}dq'
<k,q'+\frac{ls}{A}|\psi><\psi|k,q'>
e^{2\pi ij_1(\frac{q'}{l})}e^{2\pi im(\frac{lk}{2\pi})}
\end{eqnarray}       
where $q'=q-\frac{ls}{A}$. This is a calculation for the Fourier 
coefficient of periodic (in $k$ and 
$q'$) function $<k,q'+\frac{s}{A}|\psi><\psi|k,q'>\equiv f_s$, that is
\begin{equation}
f_s=\frac{A}{2\pi}\sum_{j_1 m}D_{j_1 m s}e^{-2\pi i 
j_1(\frac{q'}{l})}e^{-2\pi 
im(\frac{lk}{2\pi})}.
\end{equation}
Notice that in equation (57), $u_1^A$ and $u_2$ commute with each other. 
Thus when we obtain an explicit expression of $f_s(K,Q)$ in terms of 
$Q=e^{-2\pi i(\frac{q'}{l})}$ and $K=e^{-2\pi i(\frac{lk}{2\pi})}$ for 
real
$\frac{q'}{l}$ and $\frac{lk}{2\pi}$, we can immediately write $P$ as
\begin{equation}
P=\frac{2\pi}{A}\sum_s u_1^s f_s(u_1^A, u_2).
\end{equation}
We calculate a projector $P$ as an example in the following. Let
$|\Omega>=|0>$, which is obviously $R$ invariant. We have
\begin{eqnarray}
<kq|\psi>&=&C_{\psi}(k,q)=\frac{C_0(k,q)}{\sqrt{\frac{2\pi}{A}\sum_{n=0}^{A-1}
|C_0(k,q+\frac{ln}{A})|^2}}
\end{eqnarray}
with $C_0(k,q)=<k,q|0>$ (ref. [9]). Then the corresponding $f_s$ is
\begin{eqnarray}
f_s&=& \frac{C_0(k,q+\frac{ls}{A})C_0^*(k,q)}
{\frac{2\pi}{A}\sum_{n=0}^{A-1}C_0(k,q+\frac{ln}{A})C_0^*(k,q+\frac{ln}{A})}
\end{eqnarray}
where
\begin{eqnarray}
C_0(k,q)&=&\frac{1}{\sqrt{l}\pi^{1/4}}\theta
\left[
\begin{array}{l}
0\\
0
\end{array}
\right]
(\frac{q}{l}+\frac{\tau k}{l\tau_2},\frac{\tau}{A})
e^{-\frac{\tau}{2i\tau_2}k^2+ikq}\nonumber\\
&=&\sqrt{\frac{Ai}{l\tau\sqrt{\pi}}}\theta
\left[
\begin{array}{l}
0\\
0
\end{array}
\right]
(\frac{lk}{2\pi}+\frac{Aq}{l\tau},-\frac{A}{\tau})
e^{-\pi i\frac{Aq^2}{\tau l^2}}.
\end{eqnarray}
Thus we have for real $\frac{q}{l}=u$ and $\frac{lk}{2\pi}=v$,
\begin{eqnarray}
g(u,v)_{ss'}&=&C_0(k,q+\frac{ls}{A})C_0^*(k,q+\frac{ls'}{A})\nonumber\\
&=&\frac{1}{l\sqrt{\pi}}\theta
\left[
\begin{array}{l}
0\\
0
\end{array}
\right]
(u+\frac{s}{A}+\frac{\tau v}{A},\frac{\tau}{A})
\theta
\left[
\begin{array}{l}
0\\
0
\end{array}
\right]
(u+\frac{s'}{A}+\frac{\tau^* v}{A},\frac{-\tau^*}{A})\nonumber\\
& &e^{\pi i \frac{\tau-\tau^*}{A}v^2+2\pi i\frac{s-s'}{A}v},\\
&=&\frac{A}{l|\tau|\sqrt{\pi}}\theta
\left[
\begin{array}{l}
0\\
0
\end{array}
\right]
(v+\frac{A}{\tau}(u+\frac{s}{A}),-\frac{A}{\tau})
\theta
\left[
\begin{array}{l}
0\\
0
\end{array}
\right]
(v+\frac{A}{\tau^*}(u+\frac{s'}{A}),\frac{A}{\tau^*})\nonumber\\
& &e^{-\pi i \frac{A}{\tau}(u+\frac{s}{A})^2+\pi 
i\frac{A}{\tau^*}(u+\frac{s'}{A})^2}
\end{eqnarray}
due to 
\begin{eqnarray}
\theta
\left[
\begin{array}{l}
0\\
0
\end{array}
\right](z,\tau)^* &=& \theta
\left[
\begin{array}{l}
0\\
0
\end{array}
\right] (z*,-\tau*).
\end{eqnarray}
In the $T^2/Z_6$ case, $\tau=e^{\frac{i\pi}{3}}$, so we have
\begin{eqnarray}
g(u+1,v)&=&g(u,v+1)=g(u,v),\\
g(u+\tau,v)&=& e^{-2\pi iA(2u+\frac{v}{A}+x)}g(u,v),\\
g(u,v+A\tau)&=& e^{-2\pi i(2v+Au+y)}g(u,v) 
\end{eqnarray}
where $x=\tau-\frac{1}{2}+\frac{s+s'}{A}$ and 
$y=A\tau-\frac{A}{2}+\frac{s}{\tau}+\frac{s'}{\tau^*}$. From the above 
equation, one can prove
\begin{eqnarray}
g(u,v)&=&\{ \sum_{i=1}^{2}\theta
\left[
\begin{array}{l}
0\\
0
\end{array}
\right]
(Au+v-v_i+\frac{1}{2},\tau A)
\theta
\left[
\begin{array}{l}
0\\
0
\end{array}
\right]
(v-v_i +\frac{A\tau}{2}, \tau A)\nonumber\\
& &\theta
\left[
\begin{array}{l}
0\\
0
\end{array}
\right]
(u+\frac{s}{A}+\frac{\tau}{A}v_i,\frac{\tau}{A})
\theta
\left[
\begin{array}{l}
0\\
0
\end{array}
\right]
(u+\frac{s'}{A}+\frac{\tau^* v_i}{A},-\frac{\tau^*}{A})\nonumber\\
& &X(v_i)\}/\lbrack\theta
\left[
\begin{array}{l}
0\\
0
\end{array}
\right]
(Au+\frac{1}{2}, \tau A)
\theta
\left[
\begin{array}{l}
0\\
0
\end{array}
\right]
(\frac{\tau A}{2},\tau A)\rbrack\equiv\{G_{ss'}(u,v)\}/\lbrack r(u)
\rbrack
\end{eqnarray}                        
where
\begin{eqnarray}
v_1 &=& 
\frac{1}{2}(\frac{A}{2}+\frac{A\tau}{2}+\frac{1}{2}-\frac{As}{\tau}-
\frac{As'}{\tau^*}),\\
v_2 &=&
\frac{1}{2}(\frac{A}{2}+\frac{A\tau}{2}-\frac{1}{2}-\frac{As}{\tau}-
\frac{As'}{\tau^*}),\\
X(v_i)&=&\frac{1}{l\sqrt{\pi}}e^{-\frac{2\pi \tau_2}{A}v^2+2\pi iv(
\frac{s}{A}-\frac{s'}{A})}.
\end{eqnarray}
Thus
\begin{eqnarray}
f_s&=& \frac{G_{s0}(u, 
v)}{\frac{2\pi}{A}\sum_{n=0}^{A-1}G_{nn}(u, v)}.
\end{eqnarray}
 The proof is as follows. Because of equations (67) and (69), the entire 
function 
$g$
of $v$ belongs to a two dimensional function space of $v$. Properly choose
two functions as a base. Then fix the coefficients (they are functions of 
$u$) at two special values of $v$ ($v_1$ and $v_2$), we finally obtain 
equation (70). Let
\begin{equation}
\theta
\left[
\begin{array}{l}
0\\
0
\end{array}
\right]
(z,\tau)\equiv \Theta(Z,\tau)=\sum_me^{\pi im^2\tau}Z^m, ~~~~~~Z=e^{-2\pi 
iz}.
\end{equation}
Then
\begin{eqnarray}
G_{ss'}(u,v)&=&\sum_{i=1}^{2}\Theta(KQ^Ae^{2\pi 
i(v_i-\frac{1}{2})},A\tau)
\Theta(Ke^{2\pi i(v_i-\frac{A\tau}{2})},A\tau)\Theta(Qe^{-2\pi 
i(\frac{\tau}{A}v_i+\frac{s}{A})},\frac{\tau}{A})\nonumber\\
& &\Theta(Qe^{-2\pi
i(\frac{\tau^*}{A}v_i+\frac{s'}{A})},-\frac{\tau^*}{A})X(v_i)
\equiv \Phi_{ss'}(K,Q),
\end{eqnarray}  
giving
\begin{eqnarray}
P&=& \sum_{s=0}^{A-1} u_1^s\frac{\Phi_{s0}( 
u_1^A, u_2)}{\sum_{n=0}^{A-1}\Phi_{nn}(u_1^A, u_2)}.
\end{eqnarray}
This is an explicit $T^2/Z_6$ projector. 

\section{Complete set of projections}

We assume that all operators in $T^2$ can be expressed as
\begin{eqnarray}
\hat{B}&=& \sum_{j_1j_2}U_1^{j_1}U_2^{j_2}\hat{b}U_2^{-j_2}U_1^{-j_1}
\end{eqnarray}
for some operators $\hat{b}$ in noncommutative space $R^2$. 
Reorganize the complete set $|k,q>$ as $\{|k,q>\}=\{|k,q_0,s>\},~~q=q_0+
\frac{l}{A}s$, where $k\in [0,\frac{2\pi}{l}), q_0\in [0, \frac{l}{A}),~~
s=0,\cdots,A-1$. Equation (19) becomes
\begin{eqnarray}
id 
&=&\sum_{s=0}^{A-1}\int_{0}^{\frac{2\pi}{l}}dk\int_{0}^{\frac{l}{A}}dq_0 
|k,q_0, s><k,q_0, s|.
\end{eqnarray}
Combining the above equation and  
\begin{eqnarray}
\sum_{j}e^{ijx}&=& \sum_m2\pi \delta(x+2\pi m).
\end{eqnarray}
We can get
\begin{eqnarray}
<k,q_0+\frac{l}{A}n|\hat{B}|k',q'_0+\frac{l}{A}n'>
 &=&\frac{(2\pi)}{A}\delta(k'-k)\delta(q'_0-q_0)
<k,q_0+\frac{l}{A}n|\hat{b}|k,q_0+\frac{l}{A}n'>.
\end{eqnarray}
So we have
\begin{eqnarray}
& &<k,q_0+\frac{l}{A}n|\hat{A}\hat{B}|k',q'_0+\frac{l}{A}n'>\nonumber\\
&=&<k,q_0+\frac{l}{A}n|\hat{A}(id)\hat{B}|k',q'_0+\frac{l}{A}n'>\nonumber\\
&=&\int_0^{\frac{2\pi}{l}}\int_0^{\frac{l}{A}}\sum_{n"}
<k,q_0+\frac{l}{A}n|\hat{a}|k",q"_0+\frac{l}{A}n"><k",q"_0+\frac{l}{A}n"|
\hat{b}|k',q'_0+\frac{l}{A}n'>dk"dq"_0\nonumber\\
&=&\int_0^{\frac{2\pi}{l}}\int_0^{\frac{l}{A}}\frac{(2\pi)^2}{A^2}
\sum_{n"}\delta(k"-k)\delta(k'-k")
\delta(q'_0-q"_0)\delta(q"_0-q_0)\nonumber\\
& &<k,q_0+\frac{l}{A}n|\hat{a}|k,q_0+\frac{l}{A}n"><k",q"_0+\frac{l}{A}n"|
\hat{b}|k",q"_0+\frac{l}{A}n'>dk"dq"_0\nonumber\\ 
&=&\frac{(2\pi)^2}{A^2}\sum_{n"}\delta(k'-k)\delta(q'_0-q_0)
<k,q_0+\frac{l}{A}n|\hat{a}|k,q_0+\frac{l}{A}n"><k,q_0+\frac{l}{A}n"|
\hat{b}|k,q_0+\frac{l}{A}n'>.
\end{eqnarray}                          
From equation (81) and (82), one concludes that the necessary and 
sufficient 
condition of a projection operator in $T^2$ is
\begin{equation}
\sum_{n"}M_b(k,q_0)_{nn"}M_b(k,q_0)_{n"n'}=M_b(k,q_0)_{nn'}              
\end{equation}
at each point $(k,q_0)$, where $ 
M_b(k,q_0)_{n'n"}\equiv \frac{2\pi}{A}<k,q_0+\frac{l}{A}n'|\hat{b}
|k,q_0+\frac{l}{A}n">$,
The matrix $M(k,q_0)$ satisfying the above equation is always 
diagonizable. Thus 
\begin{eqnarray}
M_b(k,q_0)&=&S(k,q_0)^{-1}\bar{M}_b(k,q_0)S(k,q_0)
\end{eqnarray}
gives the complete set of projections of the form (78) in $T^2$, where $S$
is an arbitrary invertible matrix and $\bar{M}$ is a diagonized matrix 
with entry $0$ and $1$, that is to say
\begin{eqnarray}
\bar{M}_b(k,q_0)_{nn'}&=&\delta_{nn'}\epsilon_n(k,q_0),\\
\epsilon_n(k,q_0)&=&\{0,1\}.
\end{eqnarray}
We next study the conditions for $P$ being invariant under rotation $R$,
\begin{equation}
R^{-1}PR=P.
\end{equation}
First notice that from equation (5), the common eigenvectors ${|k, q_0+
\frac{l}{A}s>}$ of $U_1$ and $U_2$ are still eigenvectors of
\begin{eqnarray}
U'_1&=& R^{-1}U_1 R,~~~~~~~U'_2=R^{-1}U_2 R.
\end{eqnarray}
We have
\begin{eqnarray}
U'_1|k,q_0+\frac{l}{A}s>&=&\lambda_1|k,q_0+\frac{l}{A}s>,\\
U'_2|k,q_0+\frac{l}{A}s>&=&\lambda_2|k,q_0+\frac{l}{A}s>,
\end{eqnarray}
giving
\begin{eqnarray}
U_1 R|k,q_0+\frac{l}{A}s>&=&\lambda_1 R|k,q_0+\frac{l}{A}s>,\\
U_2 R|k,q_0+\frac{l}{A}s>&=&\lambda_2 R|k,q_0+\frac{l}{A}s>.
\end{eqnarray}  
Thus $R|k,q_0+\frac{l}{A}s>$ is still a common eigenvector of $U_1$ and
$U_2$. Since the eigenvalue of $U_2$ is $A$ fold degenerate, we conclude
from equation (19)
\begin{eqnarray}
R|k, q_0+\frac{l}{A}n>&=& \sum_{n'}A(k,q_0)_n^{n'}|k', q'_0+\frac{l}{A}n'>
\end{eqnarray}
for a definite $(k',q'_0)$. One can derive explicit relations for 
$(k',q'_0)$ and 
$(k, q_0)$ for all $Z_N$ cases, which is 
essentially linear relations $W: (k, q_0)\rightarrow (k', q'_0)$ and 
$(W)^N=id$. Since $R$ is unitary, the matrix $A(k,q_0)$ is also 
unitary, namely
\begin{eqnarray}
A^*(k,q_0)_{nn'}&=&A^{-1}(k,q_0)_{n'n}.
\end{eqnarray}
One can show that the map $W$ is an area preserving map, thus 
\begin{eqnarray}
\delta(k_1-k_2)\delta(q_{01}-q_{02})&=&
\delta(k'_1-k'_2)\delta(q'_{01}-q'_{02}).
\end{eqnarray}
Then we have 
\begin{eqnarray}
& &<k_1, q_{01}+\frac{l}{A}n_1|R^{-1}PR|k_2, 
q_{02}+\frac{l}{A}n_2>\nonumber\\
&=&\delta(k'_1-k'_2)\delta(q'_{01}-q'_{02})<k'_1,
q'_{01}+\frac{l}{A}n'_1|\hat{b}|k'_2, q'_{02}+\frac{l}{A}n'_2>
A^*(k_1,q_{01})_{n_1n'_1} A(k_2,q_{02})_{n_2n'_2}
\nonumber\\
&=&\delta(k_1-k_2)\delta(q_{01}-q_{02})<k1, 
q_{01}+\frac{l}{A}n_1|\hat{b}|k_2, q_{02}+\frac{l}{A}n_2>.
\end{eqnarray}
At last, we obtain
\begin{eqnarray}
M(kq_0)_{n_1n_2}&=&
\sum_{n'_1n'_2}A^{*}(k,q_0)_{n_1n'_1}M_(k'q'_0)_{n'_1n'_2}A(k,q_0)_{n_2n'_2},\\
M(k'q'_0) &=&A^{t}(k,q_0)M(kq_0)A^{*}(k,q_0).
\end{eqnarray}
That is, from the matrix $M$ of a giving point $(k,q_0)$, we can get a 
definite $M$ for the point $(k',q'_0)=W (k,q_0)$, if the corresponding 
operator is $R$ rotation invariant.
From the explicit expression of $W$, we can show that one can always 
divide the area $\sigma : (k\in [0, \frac{2\pi}{l}), q_0\in [0, 
\frac{l}{A}))$ into $N$ pieces $\sigma_1,\sigma_2,\cdots,\sigma_N$ for
$T^2/Z_N$, where $W : \sigma_i\rightarrow \sigma_{i+1}, \sigma_N 
\rightarrow \sigma_1 $. We can arbitrarily choose $M(k,q_0)$ in $\sigma_1$
by equation (84) and get $M(k,q_{0})$ in $\sigma_2,\cdots,\sigma_N$ by 
equation (98). In such way we obtain all projective operators in $T^2/Z_N$ 
by series expansion 
\begin{eqnarray}
P&=&\sum_{j_1j_2} D_{j_1j_2}e^{-\frac{2\pi 
i}{l}(j_1\hat{y_1}+\frac{j_2-\tau_1 j_1}{\tau_2}\hat{y_2})}\nonumber\\
D_{j_1j_2}&=& \frac{1}{A}tr\lbrack \hat{b} e^{\frac{2\pi 
i}{l}(j_1(\hat{y_1}-\frac{\tau_1}{\tau_2}\hat{y_2})+j_2 
\frac{\hat{y_2}}{\tau_2})}\rbrack\nonumber\\
&=&\frac{1}{A}\sum_{nn'}\int dkdq_0dk'dq'_0
<k,q_0+\frac{l}{A}n|\hat{b}|k',q'_0+\frac{l}{A}n'>\nonumber\\
& &<k',q'_0+\frac{l}{A}n'|
e^{\frac{2\pi i}{l}(j_1(\hat{y_1}-
\frac{\tau_1}{\tau_2}\hat{y_2})+j_2 \frac{\hat{y_2}}{\tau_2})}
|k,q_0+\frac{l}{A}n>\nonumber\\
&=&\frac{1}{A}\sum_{nn'}\int 
dkdq_0dk'dq'_0<k,q_0+\frac{l}{A}n|\hat{b}|k',q'_0+\frac{l}{A}n'>\nonumber\\
& &\delta(k-k')\delta(q_0-q'_0)E(k,q_0,j_1,j_2)_{n'n}\nonumber\\
&=& \sum_{nn'}\int dkdq_0 
(2\pi)^{-1}M(k,q_0)_{nn'}E(k,q_0,j_1,j_2)_{n'n}\nonumber\\
&=&\int dkdq_0 (2\pi)^{-1} tr M(k,q_0) E(k,q_0,j_1,j_2)
\end{eqnarray} 
where 
\begin{eqnarray}
E(k,q_0,j_1,j_2)_{n'n}&=& e^{2\pi ij_1(q_0+\frac{l}{A}n')/l}e^{\frac{\pi 
ij_1j_2}{A}}\nonumber\\
& &e^{i(j_2+n'-n)\frac{lk}{A}}\sum_{j}\delta_{j_2+n'-n,jA}. 
\end{eqnarray}

\end{document}